MATERIALS SCIENCE

# Manipulating surface magnetic order in iron telluride

Christopher Trainer[1], Chi M. Yim[1], Christoph Heil[2,3], Feliciano Giustino[2], Dorina Croitori[4,5], Vladimir Tsurkan[4,5], Alois Loidl[4], Efrain E. Rodriguez[6], Chris Stock[7], Peter Wahl[1]*

Control of emergent magnetic orders in correlated electron materials promises new opportunities for applications in spintronics. For their technological exploitation, it is important to understand the role of surfaces and interfaces to other materials and their impact on the emergent magnetic orders. Here, we demonstrate for iron telluride, the nonsuperconducting parent compound of the iron chalcogenide superconductors, determination and manipulation of the surface magnetic structure by low-temperature spin-polarized scanning tunneling microscopy. Iron telluride exhibits a complex structural and magnetic phase diagram as a function of interstitial iron concentration. Several theories have been put forward to explain the different magnetic orders observed in the phase diagram, which ascribe a dominant role either to interactions mediated by itinerant electrons or to local moment interactions. Through the controlled removal of surface excess iron, we can separate the influence of the excess iron from that of the change in the lattice structure.

## INTRODUCTION

Multiple competing interactions in strongly correlated electron materials lead to a plethora of emergent phases, which are highly sensitive to external stimuli and offer tremendous potential for applications. Exploiting these requires interfacing them to the outside world, yet relatively little is known about the influence of reduced symmetries and the interface itself. Iron telluride ($Fe_{1+x}Te$) is such a strongly correlated electron material with a complex magnetic phase diagram as a function of excess iron concentration $x$. At low excess iron concentration ($x < 0.11$), a bicolinear antiferromagnetic (AFM) order with an ordering wave vector $q_{AFM} = (0.5, 0, 0.5)$ (in units of the reciprocal lattice) is observed in a crystal structure with monoclinic distortion (1–3). With increasing $x$, the crystal structure becomes orthorhombic ($x > 0.11$), with a reduction in the difference in the lattice constants in the $a$ and $b$ directions (2–4). This change is accompanied by the development of patches with incommensurate magnetic order, which coexists globally with the bicolinear order for $0.11 < x < 0.14$ (5). For $x > 0.14$, the order becomes fully incommensurate, and a helimagnetic spin spiral develops (1, 2).

The bicolinear magnetic structure at low excess iron concentrations ($x < 0.11$) is well reproduced by density functional theory (DFT) calculations (6, 7), whereas accounting for the influence of interstitial iron has been more challenging. The incommensurate spiral structure can be reproduced by assuming that the interstitial Fe atoms lead to electron doping (8) or by considering additional nearest-neighbor coupling due to the randomly distributed interstitial Fe atoms (9). Even for low interstitial iron concentrations, multi-$q$ plaquette order has been predicted as a result of magnetic frustration and quantum fluctuations (10–12).

It is only very recently that real space imaging of the surface magnetic order in iron telluride has been demonstrated by spin-polarized scanning tunneling microscopy (SP-STM) (13–15). In this work, we use atomic-scale imaging by low-temperature SP-STM (16–18) to determine and manipulate the surface magnetic order in iron telluride at different excess iron concentrations. Our results enable us to assess the impact of the structural distortion and excess iron concentration on the magnetic structure. By manipulating the excess iron concentration of the surface layer, we discover a double-$q$ magnetic order, which is stabilized as the bulk crystal structure becomes orthorhombic.

Low-temperature SP-STM measurements are performed at temperature $T = 2K$ to study samples with excess Fe concentration ranging between 0.04 and 0.2. Ferromagnetic tips for SP-STM are prepared by picking up excess iron atoms from the sample surface (13, 14). This ability to manipulate the excess iron concentration in the surface layer offers the opportunity to control the surface magnetic structure and discriminate between the various scenarios for the influence of interstitial iron. The lattice constants of the surface layer remain locked to the crystal structure of the bulk material independent of the excess iron concentration in the surface layer. By removing the excess Fe atoms from the surface, we are therefore decoupling the effects of the presence of the excess Fe atoms and the orthorhombic crystal distortion on the surface magnetic order. Effectively, this allows for the determination of how the magnetic order of stoichiometric $Fe_{1+x}Te$ would change if the crystal were strained from a monoclinic to an orthorhombic structure. This method of removing Fe atoms from the surface layer of orthorhombic $Fe_{1+x}Te$ thereby provides an alternative way to engineer an effective strain in the surface layer as opposed to using externally applied strain (19).

We will first demonstrate that SP-STM characterization of $Fe_{1+x}Te$ as a function of excess iron concentration reproduces the magnetic phase diagram found by bulk characterization techniques and then show how the magnetic order changes by manipulating the excess iron concentration of the surface layer.

## RESULTS

### Surface magnetic order of $Fe_{1+x}Te$

Figure 1A shows a topographic STM image of the surface of $Fe_{1.06}Te$ obtained immediately after first approaching the sample and was recorded with a nonmagnetic PtIr tip prepared by field emission on a gold target. In the image, the square lattice corresponds to the Te atoms in the top surface layer. From large-scale STM imaging of the surface, the bright protrusions are found to have the same concentration as the excess Fe in any particular Fe$_{1+x}$Te sample studied, and hence, they are identified as the excess Fe atoms. Imaging the surface with a magnetic tip, prepared by collecting Fe atoms from the sample surface (13, 14), leads to the appearance of an additional stripe-like modulation (Fig. 1B) running along the crystal $b$ axis. This modulation has a periodicity along the crystal $a$ axis twice that of the Te lattice and results in the development of an additional peak in the Fourier transform at a wave vector of $\mathbf{q}_{AFM} = (0.5, 0)$ [relative to the reciprocal lattice vectors with $\mathbf{q}_{Te} = (1, 0), (0, 1)$ that arise from the atomic lattice; compare the Fourier image insets of Fig. 1, A and B]. This additional modulation arises from the bicollinear AFM order in Fe$_{1+x}$Te with low levels of excess Fe concentration ($x < 0.11$) (13–15). The AFM magnetic order can be seen in the STM images due to tunneling magnetoresistance: For a fixed tip-sample distance, the current depends on the relative magnetization of the tip and the sample. The images shown here are constant current images, resulting in the magnetic contrast being observed in the topographic SP-STM image. The strength of this contrast depends on the relative magnetizations of tip and sample and on the spin polarization of the charge carriers in the tip and the sample at the Fermi energy. Imaging the magnetic structure of Fe$_{1.06}$Te with the same ferromagnetic tip but with its magnetization aligned along two opposite in-plane directions results in phase reversal in the appearance of the stripe-like modulation in SP-STM images (Fig. 1, C and D) (13–15). Subtraction of these SP-STM images produces a real-space image of the magnetic structure (Fig. 1E). The height difference $\Delta z$ is proportional to the spin polarization of the tunneling current and provides information on the local magnetic order projected onto the magnetization direction of the tip. The inset in Fig. 1E shows for comparison a spin-polarized image simulated on the basis of a DFT calculation, showing excellent agreement with the experimental data (see section S1).

[1]SUPA, School of Physics and Astronomy, University of St Andrews, North Haugh, St Andrews, Fife KY16 9SS, UK. [2]Department of Materials, University of Oxford, Parks Road, Oxford OX1 3PH, UK. [3]Institute of Theoretical and Computational Physics, Graz University of Technology, NAWI Graz, 8010 Graz, Austria. [4]Center for Electronic Correlations and Magnetism, Experimental Physics V, University of Augsburg, D-86159 Augsburg, Germany. [5]Institute of Applied Physics, Academy of Sciences of Moldova, MD 2028 Chisinau, Republic of Moldova. [6]Department of Chemistry of Biochemistry, University of Maryland, College Park, MD 20742, USA. [7]School of Physics and Astronomy, University of Edinburgh, Edinburgh EH9 3JZ, UK.
*Corresponding author. Email: wahl@st-andrews.ac.uk

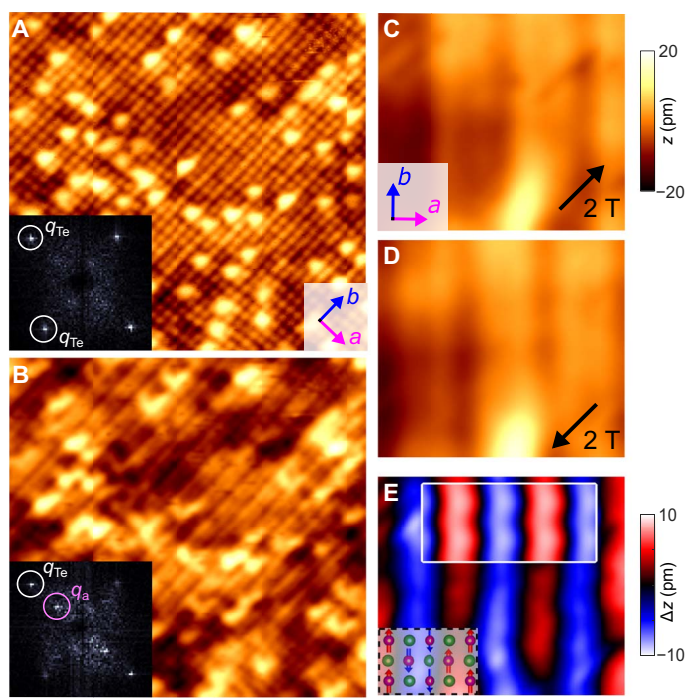

**Fig. 1. Spin-polarized STM of Fe$_{1.06}$Te.** (**A**) Topographic STM image taken with a nonmagnetic tip (14.5 by 14.5 nm$^2$). Protrusions are excess Fe atoms. Inset: Fourier transform (FT) image of (A). Peaks that arise from the Te lattice are highlighted with solid circles. (**B**) Topographic SP-STM image taken at the same position as (A) with a magnetic tip. Stripes arise from the AFM order. Inset: FT image of (B) showing additional peaks due to the AFM order. (**C** and **D**) Topographic SP-STM images taken at the same position with the tip polarized along two opposite in-plane directions (1.9 by 2.6 nm$^2$). Tunneling parameters for (A) to (D): $V = 100$ mV, $I = 50$ pA. (**E**) Difference image of (C) and (D). The height difference is proportional to the spin polarization of the tunneling current. Inset at the bottom left: Structural model of the Fe$_{1+x}$Te surface, showing the spin order in the Fe lattice (red). Inset at top right: DFT calculation of the magnetic contrast due to the spin polarization at the Fermi energy (see section S1).

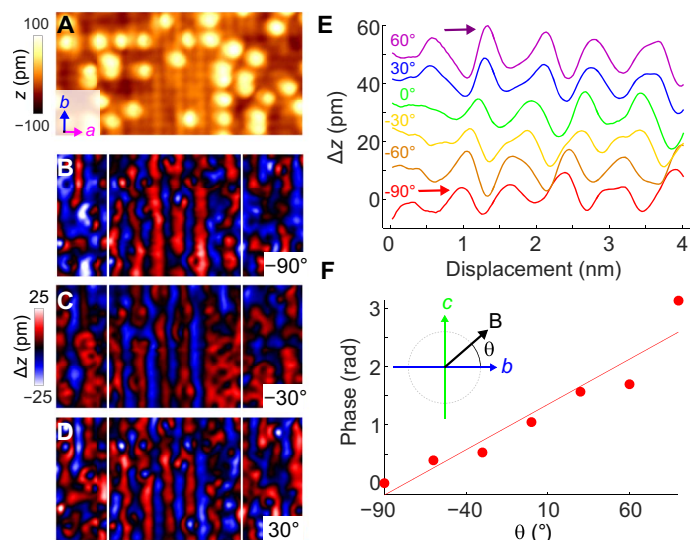

**Fig. 2. Spin spiral in Fe$_{1.16}$Te.** (**A**) Topographic SP-STM image (8.2 by 4.6 nm$^2$, $V = 50$ mV, $I = 200$ pA). (**B** to **D**) Magnetic images taken at out-of-plane angles $\theta = -90°$, $-30°$, and $30°$. Close inspection reveals that the positions of the maxima of the magnetic order shift as a function of out-of-plane angle $\theta$. (**E**) Line cuts through magnetic images as shown in (B) to (D) along $a$ for different out-of-plane field angles. The line cuts show the shift of the maxima of the stripes. (**F**) Plot of the phase of the stripes shown in (E) as a function of field angle $\theta$. The phase has been extracted using the maximum marked by an arrow in (E). Measurements were taken at an in-plane angle $\phi = 120°$ from the crystal $a$ axis.

Using the above approach and aligning the magnetization of the tip along three orthogonal directions, it becomes possible to reconstruct the surface magnetic structure of a sample (20)—provided that the sample magnetic structure remains unaltered while changing the magnetization of the tip. In the case of Fe$_{1+x}$Te, because of the high exchange interaction between the Fe atoms (15) and the considerable magnetic anisotropy energy of each Fe atom (13), the magnetic order remains unaffected by the applied fields of up to 5 T used here (14). For Fe$_{1.06}$Te, we find that the magnetic moments have an out-of-plane component, pointing into a direction $28 \pm 3°$ away from the surface plane (see section S2 and fig. S1). The magnetic ordering wave vector we find at low excess iron concentrations is in excellent agreement with previous SP-STM (13–15) and neutron scattering (1, 2) studies, as well as with calculations (6, 7). The out-of-plane angle of the magnetic order differs from that observed in neutron scattering, where the magnetization of the iron atoms is parallel to the $b$ direction but is fully consistent with previous SP-STM studies (15).

At high excess Fe concentrations $x > 0.11$, the structure of Fe$_{1+x}$Te transforms from monoclinic to orthorhombic (2, 3, 21). Figure 2A shows a topographic image of the surface of Fe$_{1.16}$Te, with three images of the magnetic order obtained with the tip magnetized in three different angles in the b-c plane (Fig. 2, B to D). As the magnetization of the tip rotates, the magnetic contrast in the difference image translates along the a direction, similar to what has been observed in previous SP-STM studies of other systems with spin spiral magnetic orders (22). This demonstrates the presence of a unidirectional spin spiral propagating along the a axis with spins rotating in the b-c plane. Analysis of the line cuts taken from the difference images (Fig. 2E) reveals the spin spiral with a wave vector of **q** = (0.43, 0), slightly incommensurate with the crystal lattice (see section S3 and fig. S2). The spin spiral found here is in full agreement with that detected in bulk samples at high excess iron concentrations $x > 0.12$ by neutron scattering.

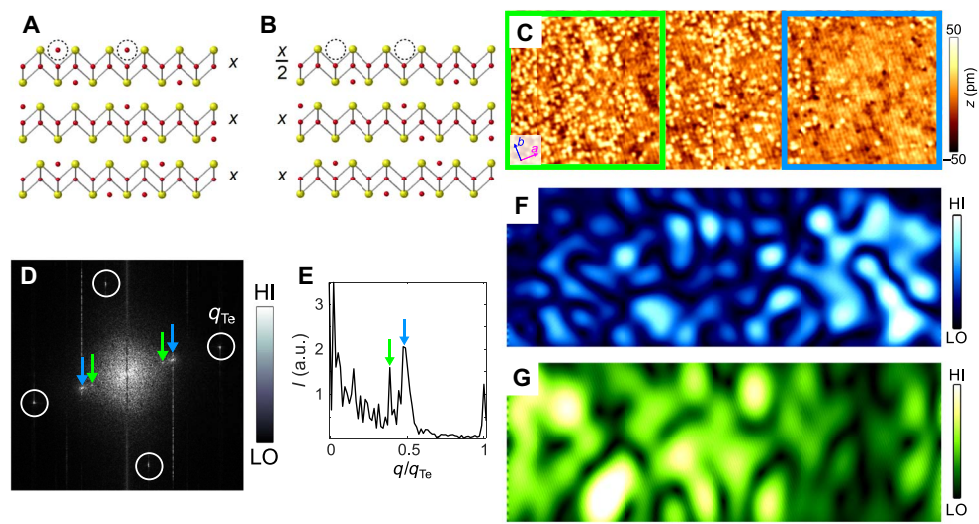

**Fig. 3. Manipulation of surface excess iron concentration.** Model of $Fe_{1+x}Te$ (**A**) before and (**B**) after removal of excess iron. Red (yellow) spheres are Fe (Te) atoms. Dashed open circles mark the interstitial/excess Fe atoms at the surface layer that are removed during surface manipulation, leading to a 50% reduction in the concentration of Fe interstitials in the surface layer (to $x/2$) compared to that of the bulk ($x$). (**C**) SP-STM image of $Fe_{1.12}Te$ (65.3 by 28.2 $nm^2$, $V$ = 150 mV, $I$ = 50 pA) showing an area where surface excess iron has been removed by the tip (blue) next to one where the excess iron has been left untouched (green). (**D**) Fourier transform of (C) showing magnetic peaks due to the bicolinear order (blue arrow) and at an incommensurate position (green arrow, see fig. S5 for Fourier transforms of regions with high and low excess iron concentrations at the surface). (**E**) Line cut from the Fourier transform in (D) taken along the $a_{Te}$ direction. Peaks corresponding to the bicolinear order at **q** = (0.5, 0) and the incommensurate order **q** = (0.39, 0) are highlighted by a blue arrow and a green arrow, respectively. (**F** and **G**) Maps of the intensity of the magnetic order at the wave vector of the bicolinear order [**q** = (0.5, 0)] and of the incommensurate order [**q** = (0.39, 0)]. Both have been obtained through Fourier filtering at the respective wave vector and then low pass filtering of the modulus. The maps show that the bicolinear order predominantly exists in regions that have been cleaned of Fe, while the incommensurate order is dominant in regions where Fe is still present. a.u., arbitrary units.

The key result from this section is that SP-STM at the surface of $Fe_{1+x}Te$ yields consistent result with neutron scattering. Differences appear merely in details, such as the out-of-plane component of the ordered moment.

### Manipulation of the excess iron concentration

In addition to imaging the surface magnetic structure at the atomic scale, STM also allows us to manipulate the surface composition, which for the case of $Fe_{1+x}Te$ can be achieved by fast, high tunneling current scanning (~2 nA) at a slow feedback loop response time. The resulting change in composition is illustrated in Fig. 3: The interstitial iron atoms originally present on the surface layer of $Fe_{1+x}Te$ are removed, leaving the remaining Fe interstitials at the lower part of the surface layer intact (for details, see section S4). As a consequence, this provides an opportunity to study the magnetic order in the surface layer with only half of the excess Fe concentration present, but with the same lattice structure as that of the bulk. Because the magnetic coupling between the layers is rather weak, as determined from inelastic neutron measurements of the spin wave excitations (23), a study of the surface magnetic structure provides us with information about how the bulk of the material would respond to similar conditions.

Figure 3C shows an SP-STM image of the surface of $Fe_{1.12}Te$, on one half of which the majority of the surface excess Fe atoms have been removed, whereas they are still present in the other half. The Fourier transform of the area shown in Fig. 3C plotted in Fig. 3D reveals two sets of magnetic ordering peaks: one at the bicolinear ordering vector **q** = (0.5, 0) and one at an incommensurate vector **q** = (0.39, 0). Fourier filtering the topographic image (Fig. 3C) at **q** = (0.5, 0) (Fig. 3F) and at **q** = (0.39, 0) (Fig. 3G), respectively, reveals that the bicolinear order is concentrated to the areas where the excess Fe atoms at the surface have been mostly removed while the incommensurate order is confined to areas where the initial surface excess Fe concentration has been left untouched. This incommensurability has also been observed in neutron scattering conducted at similar excess Fe concentrations. The STM image in Fig. 3C demonstrates that while the interstitial iron concentration of the surface layer can be manipulated using STM, the lattice structure remains commensurate with the bulk as no additional superstructure is seen, which would arise from a structural incommensurability of the surface layer with the bulk.

### Impact of excess iron concentration on surface magnetic order

A detailed investigation of the magnetic structure of the area cleaned of excess iron in $Fe_{1.12}Te$ (compare Fig. 3) reveals a complex picture: Fig. 4 (A and B) shows images of the magnetic order in the surface layer projected onto two different magnetization directions of the tip, in-plane and out-of-plane. While the image obtained with an in-plane magnetization of the tip (Fig. 4A) exhibits only the unidirectional bicolinear order, measurements with an out-of-plane magnetized tip (Fig. 4B) reveal domains of checkerboard-like double-**q** order. In the Fourier transformation (Fig. 4C), the component of the magnetic order along the $a$ axis is characterized by a sharp peak at **q** = (0.5, 0), whereas the new component along $b$ manifests itself as a broad peak at **q** ~ (0, 0.5), reflecting its localized nature. Unlike the single-**q** magnetic order found at low excess Fe concentration, which has the magnetization of the iron atoms oriented in opposite directions in the $b$-$c$ plane, in the double-**q** order, the magnetization is also modulated in the $a$ direction, leading to the formation of a spin spiral–like order. A model that reproduces this behavior is described in section S6 and equation S1. In the orthorhombic phase at higher excess iron concentration, we find an even stronger change of the magnetic structure after removing the surface interstitial iron atoms: Fig. 5A, obtained from a surface layer of Fe$_{1.1}$Te on an Fe$_{1.2}$Te sample, shows a topographic, image together with three images of the magnetic order (Fig. 5, B to D) for the tip magnetized along two in-plane (Fig. 5, B and C) and an out-of-plane direction (Fig. 5D). All three magnetic images reveal a double-**q** magnetic order, of which both components, along the $a$ and $b$ directions, are characterized by commensurate wave vectors of **q** = (0.5, 0) and (0, 0.5). As revealed by their Fourier images (insets of Fig. 5, B to D), the strength of both components varies with the magnetization of the tip. To resolve the magnetic structure, we have rotated the magnetization of the tip in the surface plane in steps of 18° through 180° and recorded a magnetic image for each angle (for details, see section S5). The results are summarized in Fig. 5E. The intensities of both components [$I(q_a)$, $I(q_b)$] vary as $|\cos\phi|$ and reach their maxima at ~30° from the $b$ axis. Both exhibit an almost identical angular dependence under the in-plane rotation. However, when the magnetization of the tip is rotated out of the surface plane, the strength of the component of the magnetic order along $a$ varies strongly, while that along $b$ remains unchanged (Fig. 5F). This can be accounted for by a model magnetic structure consisting of two spin spirals along the [110] and [1$\bar{1}$0] directions (see section S7, eq. S2, and fig. S7). The spin spirals alternate between clockwise and counterclockwise winding on alternate rows of Fe atoms.

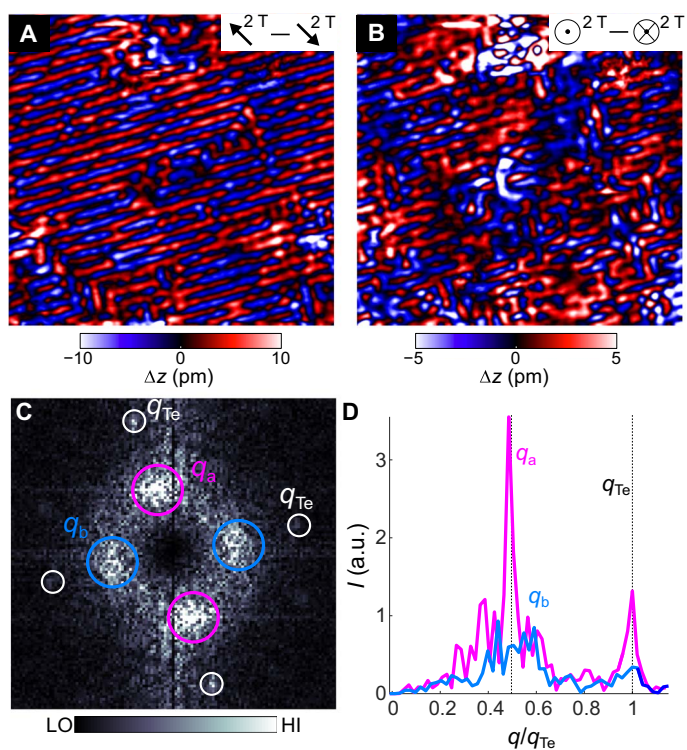

**Fig. 4. Magnetic order in a surface layer of Fe$_{1.06}$Te on Fe$_{1.12}$Te.** (**A**) Image of the magnetic order (20.5 by 20.5 nm$^2$) projected onto an in-plane direction of the magnetization as indicated by the arrows. (**B**) As in (A), taken with the same tip with an out-of-plane direction of the magnetization ($V = 100$ mV, $I = 50$ pA). (**C**) FT image of (B), with intensity at the center suppressed for clarity. Peaks due to the AFM order along $a$ and $b$ are marked with pink and cyan circles, respectively. (**D**) Normalized line cuts taken from the origin along $a$ (red) and $b$ (blue) in (C). Dashed lines indicate the positions $q_{Te}$ and $q_{AFM}$ along both $a$ and $b$ directions.

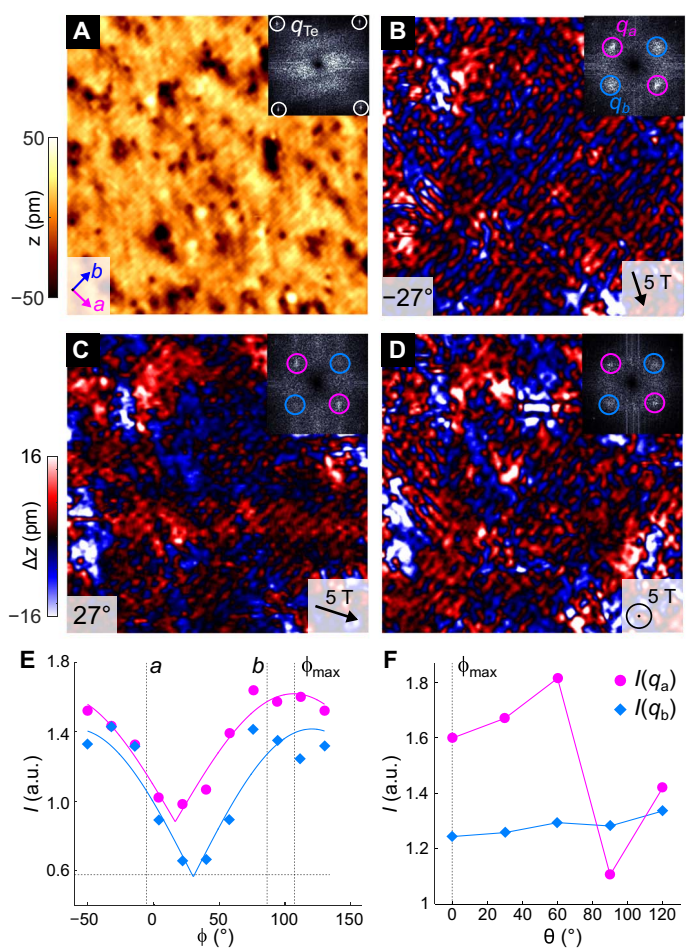

**Fig. 5. Plaquette order in on Fe$_{1.1}$Te on Fe$_{1.2}$Te.** (**A**) Nonmagnetic image (for details, see section S5 and fig. S6). Inset: FT image showing peaks due to the Te lattice. (**B**) Image of the magnetic structure for $\phi = -27°$. AFM order can be seen along both lattice directions. Inset: FT image of (B), with peaks due to AFM order in the $a$ ($b$) direction marked with pink (blue) circles. (**C** and **D**) As in (B), with (C) $\phi = 27°$ and (D) out-of-plane magnetic field ($\theta = 90°$). Insets show the corresponding FT images. Images (A) to (D) were recorded in the same area (24.1 by 24.1 nm$^2$), $V = -40$ mV, $I = 100$ pA. (**E**) Integrated intensities of the magnetic peaks in the FT image as a function of in-plane angle $\phi$. The horizontal dashed line indicates the integrated intensity of a point away from the magnetic peaks. Blue (red) markers are the intensities of the $q_b$ ($q_a$) peak. Solid lines are numerical fits of $I = I_0|\cos(\phi - \phi_0)| + c$ to the data. Vertical dashed lines indicate the in-plane field directions parallel to the $a$ and $b$ axes, as well as that of maximum intensity ($\phi = 117°$). (**F**) As in (E), plotted as a function of out-of-plane angle $\theta$ at in-plane angle of maximum intensity of the magnetic order, $\phi = 117°$. Data shown here were recorded in the same location and with the same tip.

## DISCUSSION

By combining spin-polarized imaging with the ability to manipulate the surface Fe atoms, our results enable us to extract a comprehensive picture of the magnetic phase diagram of iron telluride and assess the impact of interstitial iron on the magnetism in this material.

In Fig. 6A, we show models of the magnetic structure deduced from our study of the surface magnetic structure after removal of the interstitial iron, effectively for a surface layer of $Fe_{1+x/2}Te$ on $Fe_{1+x}Te$. In the monoclinic phase, we observe the same magnetic order after removal of interstitial iron as in the bulk. With reduced asymmetry of the lattice constants in the $a$ and $b$ directions, we see an increasing deviation of the spins from the $b$-$c$ plane. At the transition of the bulk from the monoclinic to the orthorhombic structure at $x = 0.12$, the surface layers exhibit patches of double-$q$ order, with apparent competition between the bicolinear order in the $a$ direction and a developing spin density wave (SDW) order in the $b$ direction. At even higher excess iron concentration in the bulk, the cleaned surface layer adopts a double-$q$ magnetic order. As opposed to the bulk, the surface magnetic order remains commensurate after removing the excess iron atoms.

This result is summarized in the phase diagram in Fig. 6B, where we compare the incommensurability seen in neutron scattering with the appearance of magnetic order at a second $q$-vector along $b$ in our SP-STM measurements and the anisotropy of the crystal structure. To this end, we plot the ratio $I(\mathbf{q}_b)/I(\mathbf{q}_a)$ of the intensity $I(\mathbf{q})$ of the magnetic order in the $b$ direction at $\mathbf{q}_b = (0, 0.5)$ to that in the $a$ direction, at $\mathbf{q}_a = (0.5, 0)$, as a function of excess iron concentration $x$. The doping dependence of this ratio shows that the magnetic order at $\mathbf{q}_b$ appears once the bulk has undergone the transition from a monoclinic crystal structure to an orthorhombic one.

The magnetic structure of the surface iron telluride layer adopts a staggered magnetic order in the $a$-$b$ plane, with the spins of the iron atoms alternating between fixed angles pointing away from and parallel to the $b$ axis direction, while the component along the $c$ axis alternately points out of (or into) the $a$-$b$ plane. This model of the magnetic order consists of a pair of coexisting spin spirals along the Fe-Fe [110] and [1$\bar{1}$0] directions, where the spirals alternate between right and left handedness for every other row of Fe atoms along [110]. The order remains commensurate and is similar to the plaquette order predicted theoretically in scenarios where interstitial iron is neglected and the lattice structure approaches tetragonal symmetry (10, 11) or when biquadratic exchange interactions are included (24). Djaloshinskii-Moriya interactions, which may occur for the surface layer, would promote formation of noncolinear order in the surface layer but is not expected to lead to an out-of-plane angle.

Our results suggest that a description of the interstitial iron atoms as modifying the local couplings, and thus the magnetic order, is more appropriate than there being an overall charge doping, which changes the nesting of the band structure. The latter would imply that removal of surface excess iron recovers the magnetic order found at lower excess iron concentrations, continuously changing the ordering wave vector, which, however, is not what we observe.

Our findings show some parallels to the $C_4$ magnetic order found in the iron pnictide superconductors when the magnetostructural phase transition is suppressed (25–27). In both cases, the reduction of the structural asymmetry results in formation of a double-$q$ magnetic order, although based on a different order in the undoped compound.

## CONCLUSION

Our measurements demonstrate how atomic manipulation combined with SP-STM can be used to understand the influence of defects on emergent orders (28) in strongly correlated electron materials and provide a new pathway to control emergent phases by atomic manipulation. They show that in iron telluride, it is through the monoclinic distortion that the single-**q** magnetic order is favored over a double-**q** order and that once the distortion is reduced, the magnetic structure is closer to a plaquette type of magnetic order. Our results, thus, show new opportunities to use STM-based manipulation to understand emergent magnetic orders in strongly correlated electron materials.

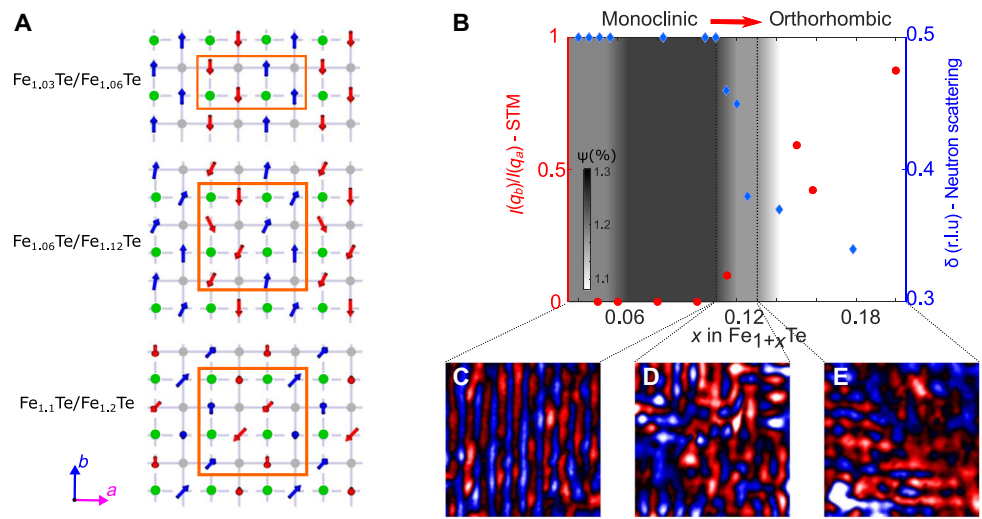

**Fig. 6. Magnetic structures and phase diagram of a surface layer of $Fe_{1+x/2}Te$.** (**A**) Model structures of surface layers of $Fe_{1.03}Te/Fe_{1.06}Te$, $Fe_{1.06}Te/Fe_{1.12}Te$, and $Fe_{1.1}Te/Fe_{1.2}Te$ that are consistent with the SP-STM results. The magnetic unit cell in each case is highlighted (for details of the model, see section S7). Arrows indicate Fe spins and are colored blue if they have a positive component along $b$ and red if they have a negative component. Green and gray spheres represent the upper and lower Te atoms. (**B**) Phase diagram of the magnetic order in the surface layer after removal of excess iron (with concentration $Fe_{1+x/2}Te$) as a function of bulk excess Fe concentration $x$. Red dots represent the ratio of the intensities $I(q_b)/I(q_a)$ of the magnetic order along the lattice directions $a$ and $b$ taken from the Fourier transforms of the STM data. The blue diamonds show the wave vector of the magnetic order in terms of the lattice spacing from neutron scattering from (2); the black dotted lines depict the "mixed spin density wave (SDW)" phase defined there. The gray background highlights how the lattice parameters change with excess Fe concentration and is defined by $\Psi = \frac{a_{Te}}{b_{Te}} - 1$, using values for $a$ and $b$ given in (2) (see also table S1). (**C** to **E**) SP-STM images (7 by 7 nm$^2$) of out-of-plane polarization of a surface layer of (C) $Fe_{1.03}Te$ in the monoclinic phase and (D) $Fe_{1.06}Te$ and (E) $Fe_{1.1}Te$ in the orthorhombic phase.

## MATERIALS AND METHODS
### STM experiments
The STM experiments were performed at 2 K using a low-temperature STM equipped with a vector magnet that enables application of magnetic fields of up to 5 T in any direction with respect to the tip-sample geometry (29). To obtain a pristine, impurity-free surface for imaging, $Fe_{1+x}Te$ samples were prepared by in situ cleaving at a temperature of ~20 K. Pt-Ir tips were conditioned by field emission on an Au(111) sample. Ferromagnetic tips used for SP-STM measurements were prepared by picking up the interstitial Fe atoms from the $Fe_{1+x}Te$ sample in the STM to create a ferromagnetic cluster of Fe atoms at the tip apex (13, 14). The magnetization of the tip was controlled through the applied magnetic field. The influence of the magnetic field on the magnetic structure of $Fe_{1+x}Te$ was assumed to be negligible due to substantial magnetocrystalline anisotropy (13). The magnetization direction of the magnetic tip is denoted by (θ, ϕ), where θ represents the out-of-plane angle and ϕ the in-plane angle measured from the crystal axes of the FeTe sample. All STM images were obtained at 2 K.

### Sample growth
Single crystals of $Fe_{1+x}Te$ were grown by the self-flux method (30, 31). The excess iron concentrations reported here were determined using energy-dispersive x-ray (EDX) analysis. Throughout the main text, the excess iron concentration of bulk samples (i.e., before removal of surface excess iron) refers to the off-stoichiometric part $x$ of the composition of the material as extracted in EDX, which, in principle, can originate either from interstitial iron or a tellurium deficiency. Interstitial or excess iron refers to iron between the FeTe layers or at the surface. Characterization of the crystals indicates that excess iron concentration and interstitial iron concentration are identical within the errors of our measurements.

## SUPPLEMENTARY MATERIALS
Supplementary material for this article is available at http://advances.sciencemag.org/cgi/content/full/5/3/eaav3478/DC1
Section S1. DFT calculation of magnetic contrast
Section S2. SP-STM study of the magnetic structure of $Fe_{1.06}Te$
Section S3. Incommensurate order in $Fe_{1.16}Te$
Section S4. Manipulating the surface excess Fe concentration
Section S5. Alternative method to determine sample spin polarization
Section S6. Model for the magnetic structure at $x = 0.12$
Section S7. Model for the magnetic structure at $x = 0.2$
Fig. S1. Spin-polarized imaging at low excess iron concentrations $x < 0.12$.
Fig. S2. Spin-polarized imaging at high excess iron concentrations $x > 0.12$.
Fig. S3. Manipulation of surface excess iron with aggressive tunneling parameters.
Fig. S4. Manipulation of surface excess iron with moderate tunneling parameters.
Fig. S5. Manipulating surface magnetic order.
Fig. S6. Extracting surface spin polarization.
Fig. S7. Simulated SP-STM images for $x = 0.12$.
Table S1. Crystal structure of $Fe_{1+x}Te$ at different excess iron concentrations $x$.
References (32–37)

**Acknowledgments**
**Funding:** C.T., C.M.Y., and P.W. acknowledge funding from EPSRC through EP/L505079/1, EP/I031014/1, and EP/R031924/1, and C.S. through EP/M01052X/1. V.T. and A.L. acknowledge funding from the Deutsche Forschungsgemeinschaft (DFG) via the Transregional Collaborative Research Center TRR 80 (Augsburg, Munich, Stuttgart). C.H. acknowledges support from the Austrian Science Fund (FWF) project no. J3806-N36 and the Vienna Science Cluster. F.G. acknowledges support from the Leverhulme Trust (grant RL-2012-001) and the U.K. EPSRC Research Council (EP/M020517/1). **Author contributions:** P.W. designed the experiment. C.T. and C.M.Y. performed the STM measurement and analyzed the data. C.H. and F.G. performed density function theory calculations. D.C., V.T., A.L., E.E.R., and C.S. grew $Fe_{1+x}Te$ single crystals. C.T. and P.W. wrote the manuscript. All authors were involved in the discussion of the manuscript. **Competing interests:** The authors declare that they have no competing interests. **Data and materials availability:** All data needed to evaluate the conclusions in the paper are present in the paper and/or the Supplementary Materials. Underpinning data will be made available at the University of St Andrews Research Portal (https://doi.org/10.17630/32f9dd7f-6749-4588-93a7-2b82218a5fdd). Additional data related to this paper may be requested from the authors.

Submitted 7 September 2018
Accepted 22 January 2019
Published 1 March 2019
10.1126/sciadv.aav3478

**Citation:** C. Trainer, C. M. Yim, C. Heil, F. Giustino, D. Croitori, V. Tsurkan, A. Loidl, E. E. Rodriguez, C. Stock, P. Wahl, Manipulating surface magnetic order in iron telluride. *Sci. Adv.* **5**, eaav3478 (2019).

# Science Advances

**Manipulating surface magnetic order in iron telluride**

Christopher Trainer, Chi M. Yim, Christoph Heil, Feliciano Giustino, Dorina Croitori, Vladimir Tsurkan, Alois Loidl, Efrain E. Rodriguez, Chris Stock and Peter Wahl